\documentclass[journal,onecolumn]{IEEEtran}
\usepackage{epsfig}
\usepackage{graphicx}
\usepackage{graphics}
\usepackage{epstopdf} 
\usepackage{tikz}
\usetikzlibrary{calc,shapes.geometric}
\usetikzlibrary{arrows,snakes,backgrounds}

\ifCLASSINFOpdf
\else
\fi

\usepackage[cmex10]{amsmath}

\def\clock{\n=\time \divide\n 60
  \m=-\n \multiply\m 60 \advance\m \time
  \ifnum \n>12 \advance\n -12 \fi
   \number\n.\twodigits\m~\ampm\time}
\def\ampm#1{\ifnum #1< 720 am\else pm\fi}
\def\twodigits#1{\ifnum #1<10 0\fi \number#1}

\usepackage{epsfig}

\def\hyptest{\renewcommand{\arraystretch}{-0.7} 
\begin{array}{c}  
\mbox{\tiny{$H_{1}$}}  \\ \vspace{-0.5 mm}
>\\ 
<\\  
\mbox{\tiny{$H_{0}$}} 
\end{array}
}

\def\nexto{\kern -0.54em}

\def\prob{{\rm {I\ \nexto P}}}

\def\pfa{{\rm P_{FA}}}

\def\Mu{\Psi}

\newcount\m \newcount\n
\def\clock{\n=\time \divide\n 60
  \m=-\n \multiply\m 60 \advance\m \time
  \ifnum \n>12 \advance\n -12 \fi
   \number\n.\twodigits\m~\ampm\time}
\def\ampm#1{\ifnum #1< 720 am\else pm\fi}
\def\twodigits#1{\ifnum #1<10 0\fi \number#1}

\begin{document}

\title{Compensating for Interference in Sliding Window Detection Processes using a Bayesian Paradigm}
\author{Graham  V. Weinberg  \\ (Draft created at \clock)\\
 }
\maketitle

\markboth{Compensating for Interference using a Bayesian Paradigm  \today}%
{}

\begin{abstract}
Sliding window detectors are non-coherent decision processes, designed in an attempt to control the probability of false alarm, for application to radar target detection. In earlier low resolution radar systems it was possible to specify such detectors quite easily, due to the Gaussian nature of clutter returns, in an X-band maritime surveillance radar context. As radar resolution improved with corresponding developments in modern technology, it became difficult to construct sliding window detectors with the constant false alarm rate property. However, over the last eight years this situation has been rectified, due to improved understanding of the way in which such detectors should be constructed. This paper examines the Bayesian approach to the construction 
of such detectors. In particular, the design of sliding window detectors, with the constant false alarm rate property, with the capacity to manage interfering targets, will be outlined.
\end{abstract}


\section{Introduction}
This paper demonstrates how the Bayesian paradigm, introduced in \cite{weinberg18} for the construction of sliding window detectors with the constant false alarm rate (CFAR) property, can be reformulated to compensate for the possible presence of an interfering target in the clutter range profile (CRP). 
It is assumed that the reader is familar with sliding window detectors and the CFAR property. The reader can consult 
 \cite{minkler90} and \cite{weinbergbook} for comprehensive overviews of the subject matter, while further examination of the Bayesian paradigm can be found in \cite{weinberg18b}.

Suppose that the statistic of the cell under test (CUT) is $Z_0$ and that the CRP is modelled by the statistics $Z_1, Z_2, \ldots, Z_N$. 
As is the usual approach, all these statistics are assumed to be independent, 
 and that they have the same common distribution function, including $Z_0$, in the absence of a target in the CUT.
The main idea is to test whether $Z_0$ is dissimilar to the clutter statistics, indicating the presence of a radar target in the CUT. 
Suppose that $H_0$ is the hypothesis that the CUT does not contain a target, and $H_1$ the alternative that the CUT contains a target embedded within clutter. One produces a single measurement of the clutter level, through a function $g(Z_1, Z_2, \ldots, Z_N)$ acting on the CRP statistics.
It is then normalised by a constant $\tau$. The test is to reject $H_0$ if and only if $Z_0$ exceeds $\tau g(Z_1, Z_2, \ldots, Z_N)$, and the presence of a target is then determined. This test can be written in the compact form
\begin{equation}
Z_0 \hyptest \tau g(Z_1, Z_2, \ldots, Z_N).\label{primarytest}
\end{equation}
The threshold multiplier $\tau$ is calculated on the basis of the expression for the probability of false alarm (Pfa), given by
\begin{equation}
\pfa = \prob(Z_0 > \tau g(Z_1, Z_2, \ldots, Z_N)| H_0), \label{primarytestpfa}
\end{equation}
where $\prob$ denotes probability. If the threshold multiplier $\tau$ can be produced such that it is independent of the clutter power, and so equivalently the Pfa is independent of the clutter power, then the test \eqref{primarytest} is said to have the CFAR property. This means that it can be used so that there is no variation in the resulting Pfa in practice. Variation of the Pfa is a disaster from a target tracking prespective \cite{weinbergbook}. 

Although the Bayesian approach has been developed for the case of Pareto Type II clutter in \cite{weinberg18}, here the principles are illustrated in the simple case of exponentially distributed clutter, corresponding to Gaussian statistics in the complex domain. It is worth observing that the basic design paradigm
will extend to more useful model for clutter, including the Pareto, Weibull, K and KK families of distributions

\section{Bayesian Paradigm for Sliding Window Detector Design}
The Bayesian paradigm views the clutter statistic's parameters as random variables and produces the Bayesian predictive density of the CUT, conditioned on the CRP, under $H_0$. This density is written  $f_{Z_0|Z_1,\ldots, Z_N}(z_0 | z_1, \ldots, z_N)$. Then the Pfa is given by
\begin{eqnarray}
\pfa &=& \prob(Z_0 > \tau | Z_1=z_1, \ldots, Z_N=z_N) 
= \int_{\tau}^\infty f_{Z_0|Z_1,\ldots, Z_N}(z_0 | z_1, \ldots, z_N) dz_0. \label{pfaexpBayes}
\end{eqnarray}
The Bayesian test is to reject $H_0$ if 
$Z_0 | Z_1=z_1, \ldots, Z_N=z_N > \tau$. The Bayesian predictive density can be derived from the likelihood function and the assumed prior distribution for the unknown parameters. As discussed in \cite{weinberg18b}, for the case of a 
one-parameter clutter model, with unknown parameter $\lambda$, the Bayesian predictive density under $H_0$ is
\begin{eqnarray}
f_{Z_0 | Z_1, \ldots, Z_N}(z_0 | z_1, z_2, \ldots, z_N) 
&=& {\int_0^\infty}{f_{Z_0 | \Lambda} (z_0 | \lambda)} f_{\Lambda | Z_1, \ldots, Z_N}
(\lambda | z_1, z_2, \ldots, z_N) f_{\Lambda}(\lambda) d\lambda, 
\label{preddens}
\end{eqnarray}
where $\Lambda$ is the random variable modelling the unknown distributional parameter, $f_{Z_0 | \Lambda} (z_0 | \lambda)$ is the density of the CUT, conditioned on $\Lambda$, $f_{\Lambda | Z_1, \ldots, Z_N}(\lambda | z_1, z_2, \ldots, z_N)$ is the density of $\Lambda$, conditioned on the CRP and 
$f_{\Lambda}(\lambda)$ is the prior distribution for $\Lambda$. 

Throughout the following the statistics of the CRP will be modelled by exponential random variables with parameter $\lambda$. It will be assumed that the target model in the CUT is Gaussian in the complex domain, so that in intensity it is also exponentialy distributed with parameter $\mu$. See \cite{gandhi88}
for further details. A simplifying assumption will be that the random variables $\Lambda$ and $\Mu$, modelling the unknown exponential distribution parameters, are independent. Realistically, $\mu$ will depend on $\lambda$; this is done to facilitate the development presented here, and to allow the specification of priors more easily.

The problem to be examined here is how to design the Bayesian decision rule, so that if one of the statistics in the CRP is an interfering target, the detector compensates for this. This analysis will proceed in three stages, beginning with the case where the interfering target's location is known.

\section{Case 1: Known Location of Interference}
Here it is assumed that there is one interfering target, which is located in the $N$th CRP cell, so that statistic $Z_N$ is a clutter plus target return.
The likelihood function is thus
\begin{equation}
f_{Z_1, Z_2, \ldots, Z_N | \Lambda, \Mu}(z_1, z_2, \ldots, z_N | \lambda, \mu) = \lambda^{N-1} e^{-\lambda \sum_{i=1}^{N-1} Z_i} 
\mu e^{-\mu Z_N}. \label{eq1}
\end{equation}
Jeffreys priors are used for both $\Lambda$ and $\Mu$, which in both cases can be shown to be the reciprocal of $\lambda$ and $\mu$ respectively.
Hence integrating the likelihood \eqref{eq1} with respect to both these improper priors, one can show the Bayesian posterior distribution of $\Lambda$ and $\Mu$, given the CRP, is
\begin{equation}
f_{\Lambda, \Mu| Z_1, Z_2, \ldots, Z_N}(\lambda, \mu) = \frac{1}{(N-2)!} \mu z_N e^{-\mu z_N} \lambda^{N-1} \left[ \sum_{i=1}^{N-1} z_i\right]^{N-1}
e^{-\lambda \sum_{i=1}^{N-1} z_i}. \label{eq2}
\end{equation}
Consequently the Bayesian predictive distribution, under $H_0$, has density
\begin{eqnarray}
f_{Z_0 | Z_1, Z_2, \ldots, Z_N}(z_0 | z_1, z_2, \ldots, z_N) &=& \int_0^\infty \int_0^\infty
\lambda e^{-\lambda z_0} z_N e^{-\mu z_N} \lambda^{N-2}
\frac{1}{(N-2)!} \left[ \sum_{i=1}^{N-1} z_i\right]^{N-1} e^{-\lambda \sum_{i=1}^{N-1} z_i} d\lambda d\mu\nonumber\\
\nonumber\\
&=& (N-1) \frac{ \left[ \sum_{i=1}^{N-1}z_i\right]^{N-1}}{ \left[ z_0 + \sum_{i=1}^{N-1}\right]^N}. \label{eq3}
\end{eqnarray}
Finally, the Pfa is obtained by integrating \eqref{eq3} from $\tau$ to $\infty$, and can be shown to be
\begin{equation}
\pfa = \left[ \frac{\sum_{i=1}^{N-1} z_i}{\tau + \sum_{i=1}^{N-1} z_i}\right]^{N-1}. \label{eq4}
\end{equation}
Hence the Bayesian test is to reject $H_0$ if $z_0 > \tau = \pfa^{-1/(N-1)} \sum_{i=1}^{N-1}z_i$, which can be recognised as the well known cell-averaging CFAR detector, with the omission of the identified interfering target. Clearly if  the interfering target is located in another CRP cell, the form of the test is much the same except this statistic is omitted. 

\section{Case 2: Unknown Location of Interference}
Under the assumption that there is an interfering target in  the CRP, it is more realistic to assume that its location is unknown. Hence suppose that the vector 
${\boldmath \pi} = [\pi_1, \pi_2, \ldots, \pi_N]$ consists of prior probabilities assumed for the event that the CRP cells contain a target.
Hence $\pi_j$ is the probability that cell $j$ contains the target. Note that $\pi_j \geq 0$ for all $j$ and $\sum_{j=1}^N \pi_j = 1$.
From a practical detector implementation, if the CUT statistic was such that in a sliding window implementation, guard cells were applied on either side of the CUT, and the CRP consisted of two sections taken from the left and the right of the guard cells, then if the interfering target was due to CUT power spread into adjacent cells, one could apply larger probabilities to such cells, and smaller probabilities to cells in the CRP further away from the CUT.

Observe that by conditional probability
\begin{equation}
\prob(Z_1 \leq z_1, Z_2 \leq z_2, \ldots, Z_N \leq z_N) 
=  \sum_{j=1}^N \prob(Z_1 \leq z_1, Z_2 \leq z_2, \ldots, Z_N \leq z_N| Z_j \mbox{ contains a target})\pi_j. \label{eq5}
\end{equation} 
One can then differentiate \eqref{eq5} to produce a density, and applying the methods of the previous section, one can show that the likelihood function is
\begin{equation}
f_{Z_1, Z_2, \ldots, Z_N | \Lambda, \Mu}(z_1, z_2, \ldots, z_N | \lambda, \mu) = \sum_{j=1}^N \lambda^{N-1} e^{-\lambda \sum_{i \not = j}z_i}
\mu e^{-\mu z_j} \pi_j. \label{eq6}
\end{equation}
It can be shown that
\begin{equation}
\int_0^\infty \int_0^\infty f_{Z_1, Z_2, \ldots, Z_N | \Lambda, \Mu}(z_1, z_2, \ldots, z_N | \lambda, \mu)  \frac{d\lambda}{\lambda} \frac{d \mu}{\mu}
= \sum_{j=1}^N \frac{\pi_j}{z_j} \frac{ (N-2)!}{\left[ \sum_{i\not = j} z_i\right]^{N-1}}, \label{eq7}
\end{equation}
and consequently the posterior distribution can be obtained through the ratio of \eqref{eq6} and \eqref{eq7}, namely
\begin{equation}
f_{\Lambda, \Mu | Z_1, Z_2, \ldots, Z_N}(\lambda, \mu) = 
\frac{\sum_{j=1}^N \lambda^{N-1} e^{-\lambda \sum_{i \not = j}z_i}
\mu e^{-\mu z_j} \pi_j}{\sum_{j=1}^N \frac{\pi_j}{z_j} \frac{ (N-2)!}{\left[ \sum_{i\not = j} z_i\right]^{N-1}}}. \label{eq8}
\end{equation}
Therefore, with an application of \eqref{eq8}, the Bayesian predictive density is
\begin{eqnarray}
f_{Z_0| Z_1, Z_2, \ldots, Z_N}(z_0 | z_1, z_2, \ldots, z_N) &=& \int_0^\infty \int_0^\infty \lambda e^{-\lambda z_0}
f_{\Lambda, \Mu | Z_1, Z_2, \ldots, Z_N}(\lambda, \mu) \frac{d\lambda}{\lambda} \frac{d\mu}{\mu} \nonumber\\
\nonumber\\
&=& \frac{ \sum_{j=1}^N \frac{\pi_j}{z_j} (N-1)! \left[ z_0 + \sum_{i\not = j}z_i\right]^{-N}}
{\sum_{j=1}^N \frac{\pi_j}{z_j} (N-2)! \left[  \sum_{i\not = j}z_i\right]^{-N+1}}. \label{eq9}
\end{eqnarray}
Finally, by integrating \eqref{eq9} as in the previous section, the Pfa can be shown to be
\begin{equation}
\pfa = \pfa(\tau) =  \frac{\sum_{j=1}^N \frac{\pi_j}{z_j} \left[ \tau + \sum_{i\not = j}z_i \right]^{-N+1}}
{\sum_{j=1}^N \frac{\pi_j}{z_j} \left[  \sum_{i\not = j}z_i \right]^{-N+1}}, \label{eq10}
\end{equation}
and since it is difficult to invert \eqref{eq10} to solve for $\tau$ one can instead reject $H_0$ if $\pfa(z_0)$ is smaller than
the design Pfa, as explained in \cite{weinberg18, weinberg18b}.

It is interesting to observe that with the choice of $\pi_N = 1$ and all other $\pi_j = 0$, the Pfa reduces to that derived in the previous section, as expected.

\section{Case 3: Uncertainty regarding Interference}
Although the detector derived in the previous section accounted for the case of uncertainty regarding the location of the interference, a shortcoming of it is that it does not include the case where there is in fact {\em no interfering target} in the CRP.  One can utilise the techniques of the previous section to compensate for this shortcoming. In this section this is outlined.

Since there is uncertainty regarding the presence of interference, redefine the vector $\pi = [\pi_0, \pi_1, \ldots, \pi_N]$, where each $\pi_j \geq 0$ and the sum of the elements of $\pi$ is unity. Here $\pi_0$ is the prior probability that there are no interfering targets, and $\pi_j$ (for $j \geq 1$) is the prior probability that there is an interfering target and it appears in CRP cell $j$.
Then \eqref{eq5} becomes
\begin{eqnarray}
\prob(Z_1 \leq z_1, Z_2 \leq z_2, \ldots, Z_N \leq z_N) 
&=&  \prob(Z_1 \leq z_1, Z_2 \leq z_2, \ldots, Z_N \leq z_N|\mbox{ no target present})\pi_0 \nonumber\\
\nonumber\\
&&
+ \sum_{j=1}^N \prob(Z_1 \leq z_1, Z_2 \leq z_2, \ldots, Z_N \leq z_N| Z_j \mbox{ contains a target})\pi_j. \label{eq11}
\end{eqnarray} 
The likelihood can then be constructed from \eqref{eq11}, and then the Bayesian posterior distribution produced as before. Once the predictive density is constructed, it can be shown that the Pfa reduces to 
\begin{equation}
\pfa(\tau) = \frac{ \pi_0 (N-1) \left[ \tau + \sum_{j=1}^N z_j\right]^{-N} + \sum_{j=1}^N \frac{\pi_j}{z_j} \left[ \tau + \sum_{i\not = j} z_i\right]^{-N+1}}
{\pi_0 (N-1) \left[\sum_{j=1}^N z_j\right]^{-N} + \sum_{j=1}^N \frac{\pi_j}{z_j} \left[ \sum_{i\not = j} z_i\right]^{-N+1}}, \label{eq12}
\end{equation}
and the test is to reject $H_0$ if and only if $\pfa(z_0)$ is smaller than the design Pfa.

It is interesting to note that the above analysis can be extended to include the case of two or more interfering targets. Additionally, the Bayesian paradigm
introduced here can also be used to design detectors robust to clutter power transitions.

\section{Conclusions}
The purpose of this paper was to indicate how the Bayesian approach could be modified to produce sliding window detectors, with the CFAR property, and with the capacity to handle interference in the CRP. The next stage of this work will be to assess performance of the detector based upon \eqref{eq12} in simulated data with and without interfering targets.

Additionally, it will be useful to apply this approach in the context of clutter modelled by Pareto statistics. A useful result, for the relevant density of the target and clutter in the interference cell, can be found in \cite{weinberg19}.

\end{document}